\documentclass[twocolumn,pra,showpacs,floatfix]{revtex4}
\usepackage{dcolumn}                 
\newcommand*{\centt}[1]{\multicolumn{1}{c}{#1}}
\newcommand*{\cent}[1]{\multicolumn{1}{c}{$#1$}}
\newcolumntype{w}[1]{D{.}{.}{#1}}
\newcommand{\br}{\vec{r}}
\newcommand{\bs}{\vec{s}}

\begin{document}

\title{Nonadiabatic corrections to the wave function and energy}

\author{Krzysztof Pachucki}
\email{krp@fuw.edu.pl}
\affiliation{Institute of Theoretical Physics,
             University of Warsaw, Ho{\.z}a 69, 00-681~Warsaw, Poland }
\author{Jacek Komasa}
\email{komasa@man.poznan.pl}
\affiliation{Faculty of Chemistry, 
        A.~Mickiewicz University, Grunwaldzka 6, 60-780~Pozna\'n, Poland }

\date{\today}

\begin{abstract}
Nonadiabatic corrections in  molecules composed of a few atoms are
considered. It is demonstrated that a systematic perturbative expansion
around the adiabatic solution is possible, with the expansion parameter
being the electron-nucleus mass ratio to the power $3/4$.  Closed form
formulae for the leading corrections to the wave function and to the energy
are derived. Their applicability is demonstrated by a comparison of
numerical results for the hydrogen molecule with the former nonadiabatic
calculations and the experimental values. Good agreement with the
recent experiment is achieved for the ground state dissociation energy of
both H$_2$ and D$_2$ .
\end{abstract}

\maketitle

\section{Introduction}

One of the most fundamental approximations of the theory of molecular
structure, the so called clamped nuclei approximation, assumes separation
of the electronic and nuclear motion and computing the electronic wave
function and energy for a fixed position of nuclei
\cite{HL27,BO27,Kol70,Kut07}. In the more accurate adiabatic approximation,
one includes the diagonal matrix element of the nuclear kinetic energy,
see Eq.~(\ref{06}). The reminder beyond this term is described as the
nonadiabatic correction.  In their calculations 
Ko\l os and Wolniewicz \cite{KW63} pointed out significance of 
nonadiabatic corrections in obtaining accurate results for H$_2$.
They claimed that within the Born-Oppenheimer (BO)
approximation  or adiabatic variants thereof, it is not possible to
estimate quantitatively the accuracy of the method, and suggested that one
should drop from the very beginning the idea of separation of the
electronic and nuclear motions, and start instead with the exact
Schr\"odinger equation for all the particles involved.  Since that work,
calculations including the nonadiabatic as well as the relativistic effects
in the hydrogen molecule have been significantly improved, among others, by
Wolniewicz and Ko{\l}os \cite{KW64,DW86,DW92,Wol95,Wol96}, Adamowicz 
\cite{KA99,BA03,KA00,SKBMA08}, Bishop \cite{BC78}, Anderson \cite{CA95} and their
coworkers.  However, no systematic theory of nonadiabatic corrections to
dissociation energies or rovibrational splittings has been pursued.  For
simple molecules such as H$_2$, direct nonadiabatic
calculations with explicitly correlated Gaussians are at present possible
\cite{KA99,KA00,BA03,SKBMA08}, but no  reliably estimate of their numerical
uncertainties has been made. In fact, results from Refs.~\cite{BA03} and
\cite{SKBMA08} are not in good agreement with each other, 
if all presented digits are significant. However, the recent
results of the variational calculation in Ref. \cite{SKBMA08} are in
remarkable agreement with the original work 
of Wolniewicz \cite{Wol95}, but our results obtained here
for total nonrelativistic energies lie 
slightly below their predictions, outside the estimated uncertainties.

At present, the lack of nonadiabatic corrections in simple molecules limits
the knowledge of physical properties, such as the dissociation energy,
vibrational, and rotational energy spacings. 
In certain instances the nonadiabatic
effects can not be considered small at all. For example, a recent study of
the spontaneous ortho-para transition in H$_2$ \cite{PK08a} reveals that
the contribution of the nonadiabatic effects to the transition probability
reaches as much as 40\%.  Another example concerns the helium dimer.  The
nonadiabatic corrections are expected \cite{jeziorsk} to significantly
contribute to the dissociation energy of the highest vibrational level
$D_{14}$ of the spin polarized helium  molecule in $^5\Sigma$ electronic
state, for which precise measurement has recently been performed \cite{ct}.
Indeed, authors of \cite{przyb} observed that using the atomic mass instead
of the nuclear mass in Eq.~(\ref{06}) \cite{Kut07} changes the dissociation
energy of the $D_{14}$ level by as much as 5\%.  This mass change 
is in accordance with the result obtained here, Eq.~(\ref{Amas}), and
indicates an importance of the nonadiabatic effects.

In this work, we construct a systematic perturbative expansion for
nonadiabatic corrections to the wave function and energy in molecules
composed of a few atoms, with the expansion parameter being the
electron-nucleus mass ratio to the power $3/4$ [$(m_{\rm e}/M)^{3/4}$]. We
derive closed form formulae for the leading nonadiabatic corrections and
present accurate numerical results for the H$_2$ and D$_2$ molecules.
This method, we suppose, can be applied to more complex molecules, where fully
nonadiabatic calculations are infeasible, which is the main advantage of
the presented here perturbative approach.

\section{Methodology}

\subsection{Theoretical framework}

Let us first consider a diatomic molecule, and assume the
reference frame be related to the mass center of the two nuclei.  The total
wave function $\phi$ is the solution of the stationary Schr\"odinger
equation 
\begin{equation}
H\,\phi = E\,\phi\,, \label{01}
\end{equation}
with the Hamiltonian 
\begin{equation}
H = H_{\rm el} + H_{\rm n}\,, \label{02}
\end{equation}
split into the electronic and nuclear parts.  In the electronic Hamiltonian
$H_{\rm el}$ 
\begin{equation}
H_{\rm el} = -\sum_{i}\frac{\nabla^2_i}{2\,m_{\rm e}} + V \label{03}
\end{equation}
the nuclei have fixed positions $\vec R_A$ and $\vec R_B$, while the
nuclear Hamiltonian is
\begin{eqnarray}\label{ETn}
H_{\rm n}&=&-\frac{\nabla^2_{\!R_A}}{2\,M_A}-\frac{\nabla^2_{\!R_B}}{2\,M_B} \\
        &=& - \frac{\nabla^2_{\!R}}{2\,\mu_{\rm n}} 
           - \frac{(\sum_{i}\,\vec\nabla_i)^2}{2\,(M_A+M_B)},\label{ETns}
\end{eqnarray}
where $\vec R = \vec R_A-\vec R_B$ and $\mu_{\rm n}$ is the nuclear reduced
mass.
 
In the adiabatic approximation the total wave function of the molecule 
\begin{equation}
\phi_{\rm a}(\vec r,\vec R) = \phi_{\rm el}(\vec r)_{\vec R} \; \chi(\vec R) \label{04}
\end{equation}
is represented as a product of the electronic wave function $\phi_{\rm el}$
and the nuclear wave function $\chi$.  The electronic wave function obeys
the clamped nuclei electronic Schr\"odinger equation 
\begin{equation}
\bigl[H_{\rm el}-E_{\rm el}(R)\bigr]\,|\phi_{\rm el}\rangle = 0, \label{05}
\end{equation} 
while the nuclear wave function obeys the Schr\"odinger equation in the
effective potential generated by electrons
\begin{equation}
\bigl[ H_{\rm n} +\bigl\langle H_{\rm n}\bigr\rangle_{\rm el}+E_{\rm el}-E_{\rm a}\bigr]\,
|\chi\rangle = 0\,. \label{06}
\end{equation} 
To account for  corrections to the adiabatic wave function $\phi_{\rm a}$
and the energy $E_{\rm a}$ resulting from the coupling between the motion
of nuclei and electrons, we formulate a systematic perturbative expansion
around the adiabatic solution, and derive explicit expression for the
leading nonadiabatic corrections to the wave function $\delta\phi_{\rm na}$
and to the energy $\delta E_{\rm na}$.

\subsection{Derivation of the nonadiabatic corrections}

The total wave function
\begin{equation}
\phi = \phi_{\rm a} + \delta\phi_{\rm na} = \phi_{\rm el}\,\chi + \delta\phi_{\rm na}
\label{07}
\end{equation}
is the sum of the adiabatic solution and a nonadiabatic correction.  Due to
the normalization of $\phi$ and $\phi_{\rm a}$, the nonadiabatic correction is
orthogonal to $\phi_{\rm a}$ in the leading order of the electron-nucleus
mass ratio
\begin{equation}
\langle\delta\phi_{\rm na}|\phi_{\rm a}\rangle = 0\,.\label{08}
\end{equation}
Let us consider the electronic matrix element
\begin{equation}
\langle\delta\phi_{\rm na}|\phi_{\rm el}\rangle_{\rm el} \equiv\delta\chi, \label{09}
\end{equation}
which is a function of the internuclear distance $R$. The nonadiabatic
correction $\delta\phi_{\rm na}$ will be  decomposed into two parts
\begin{equation}
\delta\phi_{\rm na} = \phi_{\rm el}\,\delta\chi + \delta'\phi_{\rm na}.\label{10}
\end{equation}
It follows from Eqs.~(\ref{08}) and (\ref{09}) that these two parts obey
the following orthogonality conditions
\begin{eqnarray}
\langle\delta'\phi_{\rm na}|\phi_{\rm el}\rangle_{\rm el} &=& 0\,,\label{11}\\
\langle\delta\chi|\chi\rangle &=& 0\,.\label{12}
\end{eqnarray}
Let us now revert to the stationary Schr\"odinger equation in (\ref{01})
\begin{equation}
\bigl[H_{\rm el}+H_{\rm n}-E_{\rm a}-\delta E_{\rm na}\bigr]\,
\bigl|\phi_{\rm el}\,\chi+\delta\phi_{\rm na}\bigr\rangle = 0\,,\label{13}
\end{equation}
and rewrite it to the following form
\begin{eqnarray}
\bigl[E_{\rm el}(R)-H_{\rm el}\bigr]\,|\delta\phi_{\rm na}\rangle &&
\label{14} \\ &&\hspace*{-15ex}
=\bigl[E_{\rm el}(R)+H_{\rm n}-E_{\rm a}-\delta E_{\rm na}\bigr]\,
\bigl|\phi_{\rm el}\,\chi+\delta\phi_{\rm na}\bigr\rangle.\nonumber 
\end{eqnarray}
The solution can be formally written as
\begin{eqnarray}
|\delta\phi_{\rm na}\rangle &=& |\phi_{\rm el}\,\delta\chi\rangle +
\frac{1}{(E_{\rm el}-H_{\rm el})'}
\label{15}\\ &&\times
\bigl[E_{\rm el}+H_{\rm n}-E_{\rm a}-\delta E_{\rm na}\bigr]\,
\bigl|\phi_{\rm el}\,\chi+\delta\phi_{\rm na}\bigr\rangle,\nonumber 
\end{eqnarray}
where the symbol prime in the denominator denotes exclusion of the
reference state $\phi_{\rm el}$ from the Hamiltonian inversion.  Since
$\delta\phi_{\rm na}$ is expected to be a small correction to $\phi_{\rm
a}$, it can be neglected on the right hand side in the leading order.  Then
$E_{\rm el}-E_{\rm a}-\delta E_{\rm na}$ does not contribute in the above
matrix element and this equation takes the form
\begin{equation}
|\delta\phi_{\rm na}\rangle = |\phi_{\rm el}\,\delta\chi\rangle+
\frac{1}{(E_{\rm el}-H_{\rm el})'}\,H_{\rm n}\,|\phi_{\rm el}\,\chi\rangle.\label{16}
\end{equation}
Let us point out, that the difference $E_{\rm el}-E_{\rm a}$ in
Eq.~(\ref{15}) is not necessarily small for higher rovibrational levels, so
the neglect of $\delta\phi_{\rm na}$ on the right hand side might be
questionable.  However, we will show later, that this neglect of
$\delta\phi_{\rm na}$ is in fact justified by considering the next order
nonadiabatic correction.  The presence of $E_{\rm el}$ instead of $E_{\rm
a}$ in the denominator in Eq.~(\ref{16}) is an important difference with
the former calculations of nonadiabatic corrections by Wolniewicz
\cite{Wol95}. 

The perturbed function, $\delta\chi$, is as yet unknown---we obtain it
from Eq.~(\ref{13}) by taking a matrix element with $\phi_{\rm el}$, namely
\begin{equation}
\bigl\langle\phi_{\rm el}\bigl|
E_{\rm el}+H_{\rm n}-E_{\rm a}-\delta E_{\rm na}\bigr|\,
\phi_{\rm el}\,(\chi+\delta\chi)+\delta'\phi_{\rm na}\bigr\rangle_{\rm el} = 0\label{17}
\end{equation}
The nuclear function $\chi$ obeys Eq.~(\ref{06}), which can be rewritten in
the form
\begin{equation}
\bigl\langle\phi_{\rm el}\bigl|
E_{\rm el}+H_{\rm n}-E_{\rm a}\bigr|\,\phi_{\rm el}\,\chi\bigr\rangle_{\rm el} = 0\,.
\label{18}
\end{equation}
The difference of Eqs. (\ref{17}) and (\ref{18}), neglecting the small
higher order term $\delta E_{\rm na}\,\delta\chi$ is
\begin{eqnarray}
\bigl[H_{\rm n} 
+\bigl\langle H_{\rm n} \bigr\rangle_{\rm el}+E_{\rm el}-E_{\rm a}\bigr]\,
\bigl|\delta \chi\bigr\rangle && \label{19}\\
&&\hspace{-10ex}=\delta E_{\rm na}\,\chi - 
\bigl\langle\phi_{\rm el}\bigl| H_{\rm n}\bigr|
\delta'\phi_{\rm na}\bigr\rangle_{\rm el}.\nonumber
\end{eqnarray}
If one multiplies the above equation from the left by $\chi$ and integrates
over $R$, then obtains
\begin{eqnarray}
\delta E_{\rm na} &=&  \bigl\langle\phi_{\rm el}\,\chi\bigl|
H_{\rm n}\bigr|\delta'\phi_{\rm na}\bigr\rangle
\nonumber \\ &=&
\biggl\langle\phi_{\rm el}\,\chi\biggl|
H_{\rm n}\,\frac{1}{(E_{\rm el}-H_{\rm el})'}\,
H_{\rm n}\,\biggr|\phi_{\rm el}\,\chi\biggr\rangle.
\label{Ena}
\end{eqnarray}
This is the final formula for the leading nonadiabatic correction to
energy.  At first sight, it resembles the Van Vleck sum-over-states formula
\cite{vvleck,Kol70}, but the essential difference is in that, no summation
over the nuclear states is present here, which makes Eq.~(\ref{Ena}) of
practical use.  The perturbed function $\delta\chi$ can now be inferred
from Eq.~(\ref{19}). The solution to this differential equation can be
formally written as
\begin{eqnarray}
|\delta\chi\rangle &=& \frac{1}{\bigl[E_{\rm a}- E_{\rm el}-H_{\rm n} 
-\bigl\langle H_{\rm n}\bigr\rangle_{\rm el}\bigr]'}
\bigl\langle\phi_{\rm el}\bigl|H_{\rm n}\bigr|
\delta'\phi_{\rm na}\bigr\rangle_{\rm el}.
\nonumber \\ \label{21}
\end{eqnarray}
This completes the treatment of leading nonadiabatic corrections in 
diatomic molecules. The extension of the perturbative nonadiabatic expansion to polyatomic
molecules is straightforward.  Let the total Hamiltonian be decomposed into
the electronic and nuclear parts as in Eq.~(\ref{02}), with the condition
that $H_{\rm el}$ does not contain derivatives with respect to nuclear
coordinates. Then the nonadiabatic correction to energy again takes the
form of Eq.~(\ref{Ena}).  This general formula is also valid in the case
when the $R$-independent term in $H_{\rm n}$ in Eq.~(\ref{ETns}) is shifted
to $H_{\rm el}$, which can simplify the numerical calculation of
nonadiabatic energies.

\subsection{Higher order corrections}
 
The perturbative theory presented here can be extended also to higher
orders in the electron-nucleus mass ratio in a manner similar to the
standard time independent Rayleigh-Schr{\"o}dinger perturbative expansion.
Indeed, the next order nonadiabatic correction to the energy is
\begin{eqnarray}
\delta^{(3)}E_{\rm na} &=&\biggl\langle\phi_{\rm el}\,\chi\biggl|
H_{\rm n}\,\frac{1}{(E_{\rm el}-H_{\rm el})'}\,
(H_{\rm n}+E_{\rm el}-E_{\rm a})
\nonumber \\ &&
\times\frac{1}{(E_{\rm el}-H_{\rm el})'}\,
H_{\rm n}\,\biggr|\phi_{\rm el}\,\chi\biggr\rangle
\nonumber \\ &&
+\biggl\langle\phi_{\rm el}\,\chi\biggl|
H_{\rm n}\,\frac{1}{(E_{\rm el}-H_{\rm el})'}\,
H_{\rm n}\,\biggr|\phi_{\rm el}\,\delta\chi\biggr\rangle,\label{22}
\end{eqnarray}
where $\delta \chi$ is given in Eq.~(\ref{21}), and its detailed derivation is
postponed to the next paper. Let us note, that $H_{\rm
n}+E_{\rm el}-E_{\rm a}$ in the first term of the right hand side of
Eq.~(\ref{22}) is indeed small, because $\chi$ satisfies differential
equation (\ref{06}). Therefore, the neglect of $\delta\phi_{\rm na}$ on the
right hand side of Eq.~(\ref{15}) in the derivation of the leading
nonadiabatic correction is fully justified.

The order of magnitude of the higher order corrections can be estimated by
noting that in atomic units the kinetic energy of the nuclei is of order of
the square of the nuclear displacement $\delta R $ from the equilibrium
value. It follows that $\delta R\sim (m_{\rm e}/\mu_{\rm n})^{1/4}$, and
the nuclear kinetic energy is nominally of order $\sqrt{m_{\rm e}/\mu_{\rm
n}}$.  However, it can be shown that each $H_{\rm n}$ in Eqs.~(\ref{Ena})
and (\ref{22}) contributes $(m_{\rm e}/\mu_{\rm n})^{3/4}$, therefore the
nonadiabatic correction to the energy is of order $O\bigl[(m_{\rm
e}/\mu_{\rm n})^{3/2}\bigr]$, and  the total energy can be written as 
\begin{equation}
E=E_{\rm a}+\delta E_{\rm na}\left[1+
O\biggl(\frac{m_{\rm e}}{\mu_{\rm n}}\biggr)^{3/4}\right].\label{Eorder}
\end{equation} 
This estimate may break down however, when Born-Oppenheimer potential
curves for different electronic states approach each other at certain
distance, then the denominator in Eq.~(\ref{Ena}) can be arbitrarily small.
For this situation a kind of multireference perturbation theory should be
developed and this is not a subject of the present work.

\subsection{Working equations}

The general formula (\ref{Ena}) can be readily rearranged to a more 
practical form. For this purpose, we separate out electronic matrix elements 
from the nuclear ones in $\delta E_{\rm na}$ and assume that the states $\chi$ are purely vibrational 
\begin{eqnarray}
\delta E_{\rm na} &=& \int_0^\infty\! R^2\,dR\,\left[
\chi^2\biggl\langle H_{\rm n}\,\phi_{\rm el}
\biggl|\frac{1}{(E_{\rm el}-H_{\rm el})'}\biggr| 
H_{\rm n}\,\phi_{\rm el}\biggr\rangle_{\!\rm el}\right.\nonumber \\&&
-2\,\frac{\chi\,\nabla_{\!R}^i\chi}{\mu_{\rm n}}\,
\biggl\langle H_{\rm n}\,\phi_{\rm el}
\biggl|\frac{1}{(E_{\rm el}-H_{\rm el})'}\biggr|
\nabla_{\!R}^i\phi_{\rm el}\biggr\rangle_{\!\rm el} \label{27}\\&&
+\left.\frac{\nabla_{\!R}^i\chi\,\nabla_{\!R}^j\chi}{\mu_{\rm n}^2}\,
\biggl\langle\nabla_{\!R}^i\phi_{\rm el}
\biggl|\frac{1}{(E_{\rm el}-H_{\rm el})'}\biggr|
\nabla_{\!R}^j\phi_{\rm el}\biggr\rangle_{\!\rm el}\right]\nonumber\\
&\equiv& \int_0^\infty \!R^2\,dR\,\bigl[ \chi^2(R)\,{\mathcal U}(R) 
- 2\,\chi(R)\,\chi'(R)\,{\mathcal V}(R) \nonumber \\
&&+[\chi'(R)]^2\,{\mathcal W}(R)\bigr]\,.\label{Enaw}
\end{eqnarray}
Eq.~(\ref{Enaw}) can be rewritten as an integral of the sum of two terms
which have simple physical interpretation
\begin{eqnarray}
\delta E_{\rm na} &=& \int_0^\infty \!R^2\,dR\,\biggl\{ \chi^2(R)\,
\biggl[{\mathcal U}(R) +\frac{2}{R}\,{\mathcal V}(R)+{\mathcal V}'(R)\biggr]
\nonumber \\ &&
+[\chi'(R)]^2\,{\mathcal W}(R)
\biggr\}.\label{Enaw2}
\end{eqnarray}
The first term, the coefficient of $\chi^2$, is a nonadiabatic correction to
the adiabatic energy curve, while the second term,
the coefficient of $\chi'^2$, is an $R$-dependent correction
to the nuclear reduced mass $\mu_{\rm n}$, namely
\begin{equation}
\frac{1}{2\,\mu_{\rm n}(R)} \equiv \frac{1}{2\,\mu_{\rm n}} + {\mathcal W}(R).
\label{29}
\end{equation}
An obvious advantage of the representations (\ref{Enaw}) and (\ref{Enaw2}) 
is that the electronic matrix
elements comprising the pseudopotentials $\mathcal{U}$, $\mathcal{V}$, and
$\mathcal{W}$ need to be evaluated for each $R$ only once for all
rovibrational levels. 

The electronic matrix elements in Eq.~(\ref{27}) involve 
multiple differentiation of the electronic 
wave function with respect to the internuclear distance $R$, 
which is difficult to calculate directly. 
Therefore, we rewrite these terms to a more convenient form, 
where differentiation is taken of the Hamiltonian, namely
\begin{eqnarray}
\vec \nabla_{\!R}\phi_{\rm el} &=&\frac{1}{(E_{\rm el}-H_{\rm el})'}\,
\vec\nabla_{\!R}(H_{\rm el})\,\phi_{\rm el}\,,\label{ENRf}\\
\nabla^2_{\!R}\phi_{\rm el} &\approx&\frac{1}{(E_{\rm el}-H_{\rm el})'}\,
\biggl\{\nabla_{\!R}^2(H_{\rm el})\,\phi_{\rm el}
 +2\,\vec\nabla_{\!R}(H_{\rm el}-E_{\rm el})\nonumber  \\ && \times
\frac{1}{(E_{\rm el}-H_{\rm el})'}\,
\vec\nabla_{\!R} (H_{\rm el})\,\phi_{\rm el}\biggr\},\label{ELRf}
\end{eqnarray}
where, in the last equation, a term proportional to $\phi_{\rm el}$ has been
omitted as it does not contribute to the matrix element in Eq.~(\ref{27}).

\subsection{Asymptotics} \label{ss:as}

The adiabatic  energy $E_{\rm a}$ in Eq.~(\ref{06}) and the nonadiabatic
correction $\delta E_{\rm na}$ in Eq.~(\ref{Ena}) do not vanish at large
internuclear distances. It is because they partially include the atomic
reduced mass and the so called mass polarization term. For example,
for the large atomic separation in the hydrogen molecule, 
$\langle H_\mathrm{n}\rangle_\mathrm{el}$ in Eq.~(\ref{06})
and ${\mathcal U}(R)$ in Eq.~(\ref{27}) are equal to
$m_{\rm e}/m_{\rm p}$ and $-(m_{\rm e}/m_{\rm p})^2$, accordingly,
which corresponds to first terms in the expansion of the reduced mass 
in the electron-nucleus mass ratio
\begin{equation}\label{Emas}
1-\frac{\mu}{m_{\rm e}} =\frac{m_{\rm e}/m_{\rm p}}{1+m_{\rm e}/m_{\rm p}}
= \frac{m_{\rm e}}{m_{\rm p}}
-\left(\frac{m_{\rm e}}{m_{\rm p}}\right)^2+\left(\frac{m_{\rm e}}{m_{\rm p}}\right)^3-\dots
\end{equation}
The asymptotics of ${\mathcal W}(R)$ equal to $-m_{\rm e}/m_{\rm p}^2$
corresponds to the change in Eq.~(\ref{29}) 
of the reduced nuclear mass $\mu_{\rm n}$ to the reduced atomic
mass $\mu_{\rm A}$, namely
\begin{equation}
\frac{1}{2\,\mu_{\rm A}} =
\frac{1}{2\,\mu_{\rm n}(\infty)} = \frac{1}{m_{\rm p}+m_{\rm e}} = 
\frac{1}{m_{\rm p}}\biggl(1-\frac{m_{\rm e}}{m_{\rm p}}+\cdots\biggr).
\label{Amas}
\end{equation}
Therefore, if one uses, instead of the nuclear reduced mass $\mu_{\mathrm{n}}$,
the atomic reduced mass $\mu_{\rm A}$ in the radial equation for $\chi$, 
then the large $R$ asymptotics should be subtracted from ${\mathcal W}(R)$.

\section{Calculations}\label{sec:calc}

In order to test validity of the presented perturbation theory, we consider
in more detail the hydrogen and deuterium molecules, for which the nonadiabatic
corrections were calculated perturbatively by Wolniewicz \cite{Wol95}, and
nonperturbatively by Adamowicz and coworkers \cite{KA99,KA00,BA03,SKBMA08}.
The nuclear masses used in our calculations were
$m_\mathrm{p}=1836.15267247\,m_\mathrm{e}$ for the proton and
$m_\mathrm{d}=3670.4829654\,m_\mathrm{e}$ for the deuteron, and the
conversion factor was 1 hartree = 219474.6313705 cm$^{-1}$ \cite{MT05}.

The electronic potential for the nuclear Schr{\"o}dinger
equation~(\ref{06}) consists of two contributions: 
the BO energy curve $E_{\mathrm{el}}$ and the adiabatic correction curve 
$\langle H_{\rm n}\rangle_{\rm el}$. In our calculations we used the most
accurate currently available BO potential of H$_2$ computed with subnanohartree
accuracy by Cencek \cite{Cen-pc} using 1200-term exponentially correlated 
Gaussian (ECG) wave functions \cite{RK03}. 

The adiabatic corrections for H$_2$ and D$_2$ were computed as expectation
values of the Hamiltonian $H_{\rm n}$, Eq.~(\ref{ETns}),
\begin{eqnarray}\label{Eadw}
\langle\phi_{\mathrm{el}}|H_{\rm n}|\phi_{\mathrm{el}}\rangle_{\mathrm{el}}&=&
-\frac{1}{M}\langle\phi_{\mathrm{el}}|\nabla_{\!R}^2|\phi_{\mathrm{el}}
\rangle_{\mathrm{el}} \nonumber \\
&&-\frac{1}{4M}\langle\phi_{\mathrm{el}}|
\bigl(\vec{\nabla}_1+\vec{\nabla}_2\bigr)^2|
\phi_{\mathrm{el}}\rangle_{\mathrm{el}}\,.
\end{eqnarray}
While the evaluation of the last term is straightforward, 
the first term on the right hand side is cumbersome since
it involves differentiation of the electronic wave function with 
respect to the internuclear distance.
Several methods of evaluation, either analytically \cite{KW64,BW78,LY86} or
numerically \cite{HYS86,CK97,KCR99} of the adiabatic correction $\langle
H_{\rm n}\rangle_{\rm el}$ have been reported in literature.
In this work, we note that the expectation value
$\langle\phi_{\rm el}|\nabla_{\!R}^2|\phi_{\rm el}\rangle_{\rm el}$, by
virtue of $\vec\nabla_{\!R}\langle\phi_{\rm el}|\phi_{\rm el}\rangle_{\rm
el}=0$, is equivalent to $-\langle\vec\nabla_{\!R}\phi_{\rm
el}|\vec\nabla_{\!R}\phi_{\rm el}\rangle_{\rm el}$, and with the help of 
Eq.~(\ref{ENRf}) is expressed as
\begin{eqnarray}
\langle\phi_{\mathrm{el}}|\nabla_{\!R}^2|\phi_{\mathrm{el}}\rangle_{\mathrm{el}}
&=&-\langle \phi_{\rm el}|\vec\nabla_{\!R}(H_{\rm el})\,
\frac{1}{\bigl[(E_{\rm el}-H_{\rm el})'\bigr]^2}
\nonumber\\ &&
\times\vec\nabla_{\!R}(H_{\rm el})\,|\phi_{\rm el}\rangle,
\end{eqnarray}
which can be conveniently calculated in a similar 
way as the nonadiabatic correction.

The numerical calculations were performed at a few tens of internuclear 
distances, $R$. For each $R$, three different electronic functions expanded in
an ECG basis set were involved in the evaluation of the components of the vector
$\vec{\nabla}_{\!R}\phi_{\mathrm{el}}$ of Eq.~(\ref{ENRf}).
The two-electron ECG basis functions were of the following form
\begin{eqnarray}\label{EECG}
\psi_k(\br_1,\br_2)&=&(1+\hat{P}_{12})(1+\hat\imath)\,\Xi_k\\&\times&
\exp{\left[-\sum_{i,j=1}^2 A_{ij,k}(\br_i-\bs_{i,k})(\br_j-\bs_{j,k})\right]},
\nonumber
\end{eqnarray}
where the matrices $\mathbf{A}_k$ and vectors $\bs_k$ contain nonlinear
parameters, 5 per basis function, to be optimized. The antisymmetry projector
$(1+\hat{P}_{12})$ ensures singlet symmetry, the spatial projector 
$(1+\hat\imath)$---the gerade symmetry, and $\Xi_k$ prefactor enforces
$\Sigma$ state when equal to 1, and $\Pi$ state when equal to $y_i$---the 
perpendicular Cartesian component of the electron coordinate.

The first basis, composed of 600 ECG functions (\ref{EECG}), was employed to 
expand the X$^1\Sigma_g^+$ electronic ground state wave function 
$\phi_{\mathrm{el}}$. The nonlinear parameters were optimized variationally with 
respect to the BO energy with the target accuracy of the order of nanohartree. 
Two other basis sets were employed to represent 
the two components of $\vec{\nabla}_{\!R}\phi_{\mathrm{el}}$: a 1200-term basis 
set of $^1\Sigma_g^+$ symmetry for the component parallel to the molecular 
axis and a 1200-term basis set of $^1\Pi_g$ symmetry for the perpendicular 
component.
The nonlinear parameters of these two basis sets were also optimized. This
optimization was carried out with respect to the following functional
\begin{eqnarray}
\mathcal{J}=\langle \phi_{\rm el}|\vec\nabla_{\!R}(H_{\rm el})\,
\frac{1}{(H_{\rm el}-E_{\rm el})'}\,
\vec\nabla_{\!R}(H_{\rm el})\,|\phi_{\rm el}\rangle
\end{eqnarray}
with the fixed external wave function $\phi_{\rm el}$ and
in the case of  the $\Sigma$ basis set, 
with the subtraction of the internal wave function $\phi_{\rm el}$
from the Hamiltonian inversion. This is achieved as follows. The first 600 terms
of this basis were taken from $\phi_{\mathrm{el}}$ wave function and their
nonlinear parameters were kept fixed during the optimization, only the remaining
600 terms of $\Sigma$ symmetry were optimized. This ensures that
the internal wave function $\phi_{\rm el}$ is well represented at every step
of optimization. Then the subtraction is achieved by
orthogonalization of $\vec\nabla_{\!R}(H_{\rm el})\,|\phi_{\rm el}\rangle$
with respect to the internal $|\phi_{\rm el}\rangle$. 
In the final calculations the 1200-term $\Sigma$ basis obtained according to the
above description was used also for the expansion of the external ground state 
function $\phi_{\mathrm{el}}$.

The optimized functions were employed to evaluate the adiabatic correction
(\ref{Eadw}) yielding an excellent agreement with the most accurate results
obtained by Cencek and Kutzelnigg \cite{CK97} by using the numerical 
Born-Handy method---all six digits reported in \cite{CK97} were confirmed 
by our analytic calculations. The diagonal correction function obtained this way 
was combined with the BO energy curve to form the electronic potential for the 
movement of the nuclei. The nuclear Schr{\"o}dinger
equation was solved numerically for the adiabatic energies $E_{\mathrm{a}}$
and the nuclear wave functions $\chi(R)$. 

The electronic matrix elements $\mathcal{U}$, $\mathcal{V}$, $\mathcal{W}$
entering Eq.~(\ref{Enaw}) were evaluated within the same two 1200-term ECG basis 
sets described above yielding smooth functions of $R$.
Because for the highest vibrational levels the nuclear wave functions are
spread out and contributions from larger internuclear distances are
non-negligible, the functions $\mathcal{U}(R)$, $\mathcal{V}(R)$, and
$\mathcal{W}(R)$ were represented by their asymptotic forms:
$\mathcal{U}(R)=u_0+u_6/R^6+u_8/R^8$, $\mathcal{V}(R)=v_6/R^6+v_8/R^8$, and
$\mathcal{W}(R)=w_0+w_6/R^6+w_8/R^8$, subject to
$u_0=w_0=-(m_\mathrm{e}/M)^2$ restriction (in atomic units).  
The remaining, free parameters
$u_i$, $v_i$, and $w_i$ were determined by fitting the above functions to
calculated points in the range of $(5.5:10)$ bohrs.

\begin{widetext}

\begin{table}[!ht]
\renewcommand{\arraystretch}{1.0}
\caption{Comparison of the nonadiabatic corrections to the adiabatic energy
levels of H$_2$, computed using Eq.~(\ref{Ena}), with results of Wolniewicz
\protect\cite{Wol95} as well as of Stanke~{\em et~al}. \protect\cite{SKBMA08} 
(in cm$^{-1}$).  Relative uncertainty of our results due to the neglect of higher
order corrections is 0.36\%. In the second column, our reference adiabatic energy 
$E_{\rm a}$ is given for all the vibrational levels. }
\label{TcH2}
\begin{ruledtabular}
\begin{tabular}{cw{2.11}*{3}{w{2.4}}*{2}{w{2.3}}}
\multicolumn{7}{c}{H$_2$} \\
 $v$  & \cent{E_{\rm a}[\text{au}]} & \cent{\delta E_{\rm na}} & 
 \multicolumn{1}{c}{\protect\cite{Wol95}} & \centt{Diff.}  & 
 \multicolumn{1}{c}{\protect\cite{SKBMA08}}  & \centt{Diff.} \\
\hline
 0 & -1.164\,022\,757 & -0.511 & -0.4988 & -0.012 & -0.499 & -0.012 \\
 1 & -1.145\,059\,286 & -1.347 & -1.3350 & -0.012 & -1.336 & -0.012 \\
 2 & -1.127\,168\,403 & -2.104 & -2.0913 & -0.013 & -2.092 & -0.012 \\
 3 & -1.110\,327\,841 & -2.785 & -2.7728 & -0.013 & -2.773 & -0.012 \\
 4 & -1.094\,523\,757 & -3.395 & -3.3824 & -0.013 & -3.383 & -0.012 \\
 5 & -1.079\,751\,576 & -3.934 & -3.9208 & -0.013 & -3.921 & -0.013 \\
 6 & -1.066\,017\,251 & -4.398 & -4.3847 & -0.013 & -4.385 & -0.013 \\
 7 & -1.053\,339\,044 & -4.779 & -4.7654 & -0.013 & -4.765 & -0.013 \\
 8 & -1.041\,750\,041 & -5.059 & -5.0459 & -0.013 & -5.046 & -0.013 \\
 9 & -1.031\,301\,694 & -5.210 & -5.1980 & -0.012 & -5.197 & -0.013 \\
10 & -1.022\,068\,803 & -5.187 & -5.1772 & -0.010 & -5.175 & -0.012 \\
11 & -1.014\,156\,653 & -4.926 & -4.9176 & -0.008 & -4.915 & -0.010 \\
12 & -1.007\,711\,399 & -4.333 & -4.3270 & -0.006 & -4.324 & -0.009 \\
13 & -1.002\,935\,436 & -3.287 & -3.2844 & -0.002 & -3.281 & -0.006 \\
14 & -1.000\,108\,439 & -1.646 & -1.6482 &  0.002 & -1.645 & -0.001 \\
\end{tabular}
\end{ruledtabular}

\end{table}

\end{widetext}

\section{Results and discussion}

Results of the computations are collected in Tables~\ref{TcH2} and
\ref{TcD2}, where the nonadiabatic corrections $\delta E_{\rm na}$ obtained
for all purely vibrational levels of H$_2$ and D$_2$ are listed and
compared with the existing data.  We observe small deviation from results
of Wolniewicz \cite{Wol95} for both H$_2$ and D$_2$.  
Apart from the highest vibrational levels, 
the deviation is  about $-0.013$~cm$^{-1}$ for H$_2$ and 
$-0.003$~cm$^{-1}$ for D$_2$. It can be attributed to the slightly different 
perturbation approach of Wolniewicz in Ref.~\cite{Wol95}.
Since, the fully nonadiabatic calculation by Stanke~{\em et~al}. 
in Ref.~\cite{SKBMA08} is in remarkable agreement with results of 
Ref.~\cite{Wol95} for H$_2$, the most probable explanation 
of this deviation is the neglect of the probably significant
second term in the higher correction of Eq. (\ref{22}). 
We note, that with the variational approach the obtained results are upper bounds to the 
nonadiabatic energies and the estimate of uncertainty is not necessarily
simple, as  shows the comparison with the former calculations
by Bubin and Adamowicz in \cite{BA03}. While we cannot judge
here which result is more accurate, we observe, after inclusion of
relativistic and QED corrections \cite{LJ08}, a very good agreement of our
final values with the precise experimental result for the ground state
dissociation energy in both, H$_2$, see Table~\ref{TdisH2}, and
D$_2$ molecule in Table~\ref{TdisD2}.

\begin{table}[!htb]
\renewcommand{\arraystretch}{1.0}
\caption{Comparison of the nonadiabatic corrections to the adiabatic energy
levels of D$_2$, computed using Eq.~(\ref{Ena}), with results of Wolniewicz
\protect\cite{Wol95} (in cm$^{-1}$). Relative uncertainty of our results
due to the neglect of higher order corrections is 0.21\%. In the second column, 
our reference adiabatic energy $E_{\rm a}$ is given for all the vibrational 
levels.   }
\label{TcD2}
\begin{ruledtabular}
\begin{tabular}{cw{2.11}w{2.4}w{2.4}w{2.3}}
\multicolumn{5}{c}{D$_2$} \\
 $v$ & \cent{E_{\rm a}[\text{au}]}  & \cent{\delta E_{\rm na}} & 
\multicolumn{1}{c}{\protect\cite{Wol95}} &
 \centt{Diff.}  \\
\hline
 0 & -1.1671680230 & -0.176 & -0.1725 & -0.003 \\
 1 & -1.1535267225 & -0.480 & -0.4769 & -0.003 \\
 2 & -1.1404282296 & -0.764 & -0.7605 & -0.003 \\
 3 & -1.1278630079 & -1.028 & -1.0246 & -0.003 \\
 4 & -1.1158235203 & -1.273 & -1.2698 & -0.003 \\
 5 & -1.1043043280 & -1.500 & -1.4970 & -0.003 \\
 6 & -1.0933022295 & -1.710 & -1.7062 & -0.004 \\
 7 & -1.0828164445 & -1.901 & -1.8977 & -0.004 \\
 8 & -1.0728488533 & -2.075 & -2.0709 & -0.004 \\
 9 & -1.0634043039 & -2.228 & -2.2246 & -0.004 \\
10 & -1.0544910050 & -2.361 & -2.3568 & -0.004 \\
11 & -1.0461210239 & -2.468 & -2.4646 & -0.004 \\
12 & -1.0383109190 & -2.547 & -2.5434 & -0.004 \\
13 & -1.0310825465 & -2.590 & -2.5869 & -0.004 \\
14 & -1.0244640945 & -2.590 & -2.5868 & -0.003 \\
15 & -1.0184914179 & -2.534 & -2.5316 & -0.003 \\
16 & -1.0132097815 & -2.410 & -2.4071 & -0.002 \\
17 & -1.0086761569 & -2.197 & -2.1951 & -0.002 \\
18 & -1.0049622689 & -1.875 & -1.8742 & -0.001 \\
19 & -1.0021585658 & -1.420 & -1.4196 &  0.000 \\
20 & -1.0003781627 & -0.805 & -0.8055 &  0.000 \\
21 & -0.9997348811 & -0.079 & -0.0860 &  0.007 \\
\end{tabular}
\end{ruledtabular}

\end{table}

\subsection{Simplified approach}

In the case of the lowest rovibrational states, the nonadiabatic correction
of Eq.~(\ref{27}) can be significantly simplified.  Since the derivatives
of $\phi_{\rm el}$ are here much smaller than that of $\chi$, the
contribution from terms proportional to $\chi\,\chi'$ and $\chi^2$ can be
neglected. Moreover, the nuclear wave function $\chi$ is strongly localized
around $R_0$, the average distance between nuclei, so the electronic matrix
element can be taken just at $R_0$.  The resulting simplified form of
nonadiabatic correction to energy for low lying  vibrational states is
\begin{eqnarray}\label{Eap1}
\delta E_{\rm na} &\approx&\,\frac{1}{\mu_{\rm n}^2}\,
\bigl\langle\chi'|\chi'\bigr\rangle
\biggl\langle\phi_{\rm el}\biggl|\vec n\cdot\vec\nabla_{\!R} (H_{\rm el})
\nonumber \\ &&
\frac{1}{\bigl[(E_{\rm el}-H_{\rm el})'\bigr]^3}\,
\vec n\cdot\vec\nabla_{\!R}(H_{\rm el})\biggr|\phi_{\rm el}\biggr\rangle_{R_0},
\end{eqnarray} 
where
\begin{equation}\label{Eap2}
\vec n\cdot\vec\nabla_{\!R} (H_{\rm el}) = 
\frac{\vec n}{2}\cdot\biggl(
-\frac{\vec r_{1A}}{r_{1A}^3}+\frac{\vec r_{1B}}{r_{1B}^3}
-\frac{\vec r_{2A}}{r_{2A}^3}+\frac{\vec r_{2B}}{r_{2B}^3}
\biggr)-\frac{1}{R^2}\,,
\end{equation}
and $\vec n = \vec R/R$.  The corresponding result, obtained for the ground
vibrational state of H$_2$, $-0.453$ cm$^{-1}$ differs from the accurate
one (with the asymptotics subtracted out) $-0.4458$ cm$^{-1}$ only by
1.5\%.  Moreover, this approximation can be applied also to heavier
molecules, where its accuracy should be even higher. Indeed, the simplified
calculations for the ground state of D$_2$ yield $-0.161$ cm$^{-1}$ which
differs from the accurate result $-0.1594$ cm$^{-1}$ by as few as 1.0\%.

\newcommand{\ph}{\phantom{BO}}
\begin{table}[!hb]
\renewcommand{\arraystretch}{1.0}
\caption{Dissociation energy of H$_2$ $v=0$ state (in cm$^{-1}$). Total
uncertainty is due to  higher order nonadiabatic corrections (0.0019
cm$^{-1}$) and approximate $\alpha^4$ QED correction (0.0008 cm$^{-1}$).  }
\label{TdisH2}
\begin{ruledtabular}
\begin{tabular}{lw{2.4}w{5.7}}
&\multicolumn{1}{c}{Correction} & \multicolumn{1}{c@{}}{Subtotal\qquad\qquad} \\ 
\hline
BO                         &        & 36112.593        \\
\ph + ad. corr. + asymp. (\ref{ss:as})  &        & 36118.364	\\
\ph + nonad. corr., Eq.~(\ref{Ena})     & 0.5109 & 36118.875	\\
\ph + asymp. (\ref{ss:as})              &-0.0651 & 36118.810^{a}	\\
\ph + $\alpha^2$ relat. corr. \cite{LJ08}&-0.5178& 36118.292	\\
\ph + $\alpha^3$ QED corr. \cite{LJ08}  &-0.2359 & 36118.056^{b}	\\
\ph + $\alpha^4$ QED corr. \cite{LJ08}  &-0.0017 & 36118.054	\\
uncertainty                           &\pm 0.0027& 36118.054(3)	\\
\\
Experiment \cite{ZCKSE04}               &        & 36118.062(10) \\
difference                              &        &     0.008(13) \\
\end{tabular}
$^a${\footnotesize Bubin and Adamowicz \cite{BA03}: 36118.798\hspace*{\fill}} \\
$^b${\footnotesize Wolniewicz \cite{Wol95} (includes approximate QED
corr.): 36118.069}\hspace*{\fill} 
\end{ruledtabular}

\end{table}

\begin{table}[!hbt]
\renewcommand{\arraystretch}{1.0}
\caption{Dissociation energy of D$_2$ $v=0$ state (in cm$^{-1}$). Total
uncertainty is due to  higher order nonadiabatic corrections (0.0004
cm$^{-1}$) and approximate $\alpha^4$ QED correction (0.0008 cm$^{-1}$).  }
\label{TdisD2}
\begin{ruledtabular}
\begin{tabular}{lw{2.4}w{5.7}}
&\multicolumn{1}{c}{Correction} & \multicolumn{1}{c@{}}{Subtotal\qquad\qquad} \\ 
\hline
BO                                        &         & 36746.162   \\
\ph + adiab. corr. + asymp. (\ref{ss:as}) &         & 36748.935	\\
\ph + nonad. corr., Eq.~(\ref{Ena})       &  0.1757 & 36749.111	\\
\ph + asymp. (\ref{ss:as})                & -0.0163 & 36749.094^{a}	\\
\ph + $\alpha^2$ relat. corr. \cite{LJ08} & -0.5178 & 36748.577	\\
\ph + $\alpha^3$ QED corr. \cite{LJ08}    & -0.2359 & 36748.341^{b}	\\
\ph + $\alpha^4$ QED corr. \cite{LJ08}    & -0.0017 & 36748.339	\\
uncertainty                              &\pm 0.0012& 36748.339(1)	\\
\\
Experiment \cite{ZCKSE04}                 &         & 36748.343(10) \\
difference                                &         &     0.004(11)
\end{tabular}
$^a${\footnotesize Bubin and Adamowicz \cite{BA03}: 36749.085\hspace*{\fill}} \\
$^b${\footnotesize Wolniewicz \cite{Wol95} (includes approximate QED
corr.): 36748.364\hspace*{\fill}} 
\end{ruledtabular}

\end{table}

\section{Summary}

We have formulated the perturbative approach to the calculation of
nonadiabatic effects. The leading order nonadiabatic correction to the
energy, given by Eq.~(\ref{Ena}), is conveniently  rewritten in terms of
three state independent pseudopotentials introduced in Eq.~(\ref{Enaw}),
and calculated in Sec.~\ref{sec:calc} for all vibrational levels of H$_2$ and
D$_2$. Our numerical results are slightly below, and thus in small disagreement
with calculations by Wolniewicz \cite{Wol95} and by 
Stanke {\em et al}. \cite{SKBMA08}.

Several important advantages of the approach presented here
are worth emphasizing. The first one is the possibility
to systematically derive  higher order nonadiabatic 
corrections and to control its accuracy. 
Secondly, the result for the leading 
correction of Eq.~(\ref{Ena}) can conveniently be applied 
to more complex molecules.
Thirdly, for the lowest vibrational states one can apply the approximate 
Eq.~(\ref{Eap1}), which is accurate to about 99\%, even for very light molecules 
such as H$_2$ and D$_2$. Finally, the nonadiabatic corrections to the wave
function Eq.~(\ref{16}) significantly contribute to radiative transition 
rates, such as ortho-para transition in H$_2$ \cite{PK08a}, 
or make the transition possible, as in the case of rovibrational electric dipole
transitions in HD.

\section*{Acknowledgments}
Authors wish to acknowledge inspiration by Bogumi\l\ Jeziorski. 
Part of the computations has been performed in Pozna\'n\ Supercomputing
and Networking Center.


\begin{thebibliography}{99}
\bibitem{HL27} W.~Heitler and F.~London, Z.~Phys. {\bf 44}, 455 (1927).
\bibitem{BO27} M. Born and R. Oppenheimer, 
		Ann. Phys. (Leipzig) {\bf 84}, 457 (1927).
\bibitem{Kol70} W. Ko\l os, Adv. Quantum Chem. {\bf 5}, 99 (1970).
\bibitem{Kut07} W.~Kutzelnigg, Mol. Phys. {\bf 105}, 2627 (2007).
\bibitem{KW63} W. Ko\l os and L. Wolniewicz, 
		Rev. Mod. Phys. {\bf 35}, 473 (1963).
\bibitem{KW64} W. Ko\l os and L. Wolniewicz, 
		J. Chem. Phys. {\bf 41}, 3663 (1964).
\bibitem{Wol95} L. Wolniewicz, J. Chem. Phys. {\bf 103}, 1792 (1995).
\bibitem{Wol96} L. Wolniewicz, J. Chem. Phys. {\bf 105}, 10691 (1996).
\bibitem{DW86} K. Dressler and L. Wolniewicz, 
		J. Chem. Phys. {\bf 85}, 2821 (1986).  
\bibitem{DW92} L. Wolniewicz and K. Dressler, 
		J. Chem. Phys. {\bf 96}, 6053 (1992).
\bibitem{KA99} D. B. Kinghorn and L. Adamowicz, 
		Phys. Rev. Lett. {\bf 83}, 2541 (1999).
\bibitem{BA03} S.~Bubin and L.~Adamowicz, 
		J. Chem. Phys. {\bf 118}, 3079 (2003). 
\bibitem{KA00} D. B. Kinghorn and L. Adamowicz, 
		J. Chem. Phys. {\bf 113}, 4203 (2000).
\bibitem{SKBMA08} M.~Stanke, D.~K\c edziera, S.~Bubin, M.~Molski, and L.~Adamowicz,
		J.~Chem.~Phys. {\bf 128}, 114313 (2008). 
\bibitem{BC78} D.~M.~Bishop and L.~M.~Cheung, Phys.~Rev. A {\bf18}, 1846 (1978).
\bibitem{CA95} B.~~Chen and J.~B.~Anderson, 
		J. Chem. Phys. {\bf 102}, 2802 (1995). 
\bibitem{PK08a} K. Pachucki and J. Komasa, 
		Phys. Rev. A {\bf 77}, 030501(R) (2008).
\bibitem{jeziorsk} B. Jeziorski, {\em private communication}.
\bibitem{ct} S. Moal, M. Portier, J. Kim, J. Dugu\'e, U. D. Rapol, 
  M. Leduc, and C. Cohen-Tannoudji, Phys. Rev. Lett. {\bf 96}, 023203 (2006).
\bibitem{przyb} M. Przybytek and B. Jeziorski, 
		J. Chem. Phys. {\bf 123}, 134315 (2005). 
\bibitem{vvleck} J. H. Van Vleck, J. Chem. Phys. {\bf 4}, 327 (1936).
\bibitem{MT05} P. Mohr and B.N. Taylor, Rev. Mod. Phys. {\bf 77}, 1 (2005).
\bibitem{Cen-pc} W. Cencek, {\em private communication}.
\bibitem{RK03} J. Rychlewski and J. Komasa, in Explicitly Correlated Wave
Functions in Chemistry and Physics, edited by J. Rychlewski
(Kluwer Academic Publishers, Dordrecht, 2003), p. 91-147.
\bibitem{BW78} M.~Bardo and M.~Wolfsberg, J.~Chem.~Phys. {\bf 68}, 2686 (1978).
\bibitem{LY86} B.~H.~Lengsfield and D.~R.~Yarkony, 
		J.~Chem.~Phys. {\bf 84}, 348 (1986).
\bibitem{HYS86} N.~C.~Handy, Y.~Yamaguchi, and H.~F.~Schaefer III, 
		J.~Chem.~Phys. {\bf 84}, 4481 (1986).
\bibitem{KCR99} J.~Komasa, W.~Cencek, and J.~Rychlewski, 
		Chem.~Phys.~Lett. {\bf 304}, 293 (1999).
\bibitem{CK97} W.~Cencek and W.~Kutzelnigg, 
		Chem.~Phys.~Lett. {\bf 266}, 383 (1997).
\bibitem{LJ08} G. \L ach, PhD dissertation, University of Warsaw (2008).
\bibitem{ZCKSE04} Y.~P.~~Zhang, C.~H.~Cheng, J.~T.~Kim, J.~Stanojevic, 
                  and E.~E.~Eyler, Phys.~Rev.~Lett. {\bf 92}, 203003 (2004).
\end{thebibliography}
\end{document}